\documentclass[twocolumn,showpacs,preprintnumbers]{revtex4}
\usepackage{amssymb}
\usepackage{amsmath}
\usepackage{graphicx}
\usepackage{dcolumn}
\usepackage{bm}
\usepackage{subfigure}
\usepackage{xcolor}
\usepackage[colorlinks,citecolor=blue, linkcolor=blue,hyperindex,dvipdfm]{hyperref}

\begin{document}

\title{Measurement-induced non-locality in arbitrary dimensions in terms of
the inverse approximate joint diagonalization}
\author{Li-qiang Zhang}
\author{Ting-ting Ma}
\author{Chang-shui Yu}
\email{ycs@dlut.edu.cn}
\affiliation{School of Physics, Dalian University of Technology, Dalian 116024, China }

\begin{abstract}
The computability of the quantifier of a given quantum resource is the
essential challenge in the resource theory and the inevitable bottleneck for
its application. Here we focus on the measurement-induced non-locality
and present a new definition in terms of the skew information subject to a
broken observable. It is shown that the new quantity possesses an obvious
operational meaning, can tackle the non-contractivity of the measurement-induced non-locality and has an
analytic expressions for pure states, ($2\otimes d$)-dimensional quantum
states and some particular high-dimensional quantum states. Most
importantly, an inverse approximate joint diagonalization algorithm,
due to its simplicity, high efficiency, stability and state independence, is
presented to provide the almost analytic expressions for any quantum state,
which can also shed new light on the other aspects in physics. As
applications as well as the demonstration of the validity of the algorithm,
we compare the analytic and the numerical expressions of various examples
and show the perfect consistency.
\end{abstract}

\pacs{03.67.Mn, 03.65.Ud, 03.65.Ta}
\maketitle

\section{Introduction}

Quantum mechanical phenomena such as entanglement are not only fundamental
properties of quantum mechanics, but also the important physical resources
due to their exploitation in quantum information processing tasks. Thus a
mathematical quantitative theory, i.e., the resource theory (RT), is
required to characterize the resource feature of these quantum phenomena. In
the recent years, the RT has been deeply developed for entanglement \cite%
{B1,B2,B3,o1,o2,o3}, quantum discord \cite{d1,d2,d3,d4,d5,d6} and quantum
coherence \cite{c1,c2,c3,c4,c5,c6} and so on \cite{t1,t2}. However, as the
core in RT, a good measure for the given resource of a general (especially
high-dimensional) quantum state is only available for quantum coherence \cite%
{c1,yuc}. In contrast, quantum entanglement and quantum discord can be only
well quantified in quite limited states \cite%
{peres,wootters,vidal,geom1,geom,dak,giro,diagon}. Undoubtedly, their common
challenge is the computability of the given measure (despite that some
conceptual question, e.g., the entanglement measure of multipartite states
etc., remains open). In order to study these quantum features, especially
entanglement in a general (high-dimensional) quantum system, people usually
have to seek for the lower or upper bounds for the rough reference \cite%
{o2,florian,kai,ycc}. Comparably, an effective numerical way to evaluating a
measure could be even more important for its potential applications in both
the measure itself and the large-scale quantum system. For example, the
conjugate-gradient method was proposed to calculate the bipartite
entanglement of formation \cite{nc}; the semi-definite programming method
was used to find the best separable approximation \cite{semi}; some
numerical algorithms were developed for the evaluation of the
relative-entropy entanglement of bipartite states \cite{nr,nume,compu};
3-tangle can also be evaluated by various numerical ways such as the Monto
Carlo methods \cite{nm}, the conjugate-gradient method and so on \cite%
{compu,highly,numerical}; the local quantum uncertainty can be stably
computed based on the approximate joint diagonalization algorithm \cite%
{diagon}; recently, the numerical algorithms have also been used to
calculate the quantum coherence measures \cite{robu,asym,trac,gene}.
Generally speaking, a numerical algorithm depends not only on the question
(the measure) but also on the state to be evaluated. In this sense, how to
develop an effective algorithm for a particular quantum resource especially
independent of the state is still of great practical significance.

In this paper, we present an effective numerical algorithm for the
measurement-induced non-locality \cite{loca}. the measurement-induced non-locality, dual to the quantum
discord (to some extent) \cite{d1,d2}, is one type of non-locality that
characterizes the global disturbance on a composite state caused by the
local non-disturbing measurement on one subsystem. However, the measurement-induced non-locality, similar
to the geometric quantum discord \cite{dak,geometr}, is originally defined
based on the $l_2$ norm and inherits its non-contractivity, so it is not a
good measure due to the some unphysical phenomena that could be induced \cite%
{piani}. Although the measurement-induced non-locality has been studied in many aspects \cite%
{SY,gamy,Rana,Wu,Hu,maximal,based}, no obvious operational meaning has been
found up to now and in particular, it can only be analytically calculated in
low-dimensional quantum systems \cite{loca,Wu,Hu}. Here we will first
redefine the measurement-induced non-locality based on the skew information \cite{skew,convex,uncer} in
terms of the broken observable (a complete set of rank-one projectors)
instead of a single observable. Thus not only the non-contractivity can be
automatically solved, but also an obvious operational meaning related to
quantum metrology has been found. Another distinct advantage of such a
re-definition is that it can be analytically calculated for a large class of
high-dimensional states and especially can induce a powerful numerical
means, i.e., the inverse approximate joint diagonalization, to find
the almost analytic expression for any quantum state. It is shown that this
algorithm can be stably and simply performed with even the machine precision
and especially it is state independent. The effectiveness of the algorithm
is demonstrated by comparison with the analytic results of various examples.
The remaining of this paper is organized as follows. In Sec. II, we briefly
introduce the measurement-induced non-locality and present its new definition as well as its operational
meaning. In Sec. III, we detailedly introduce the inverse approximate joint diagonalization algorithm and show
how we convert the measurement-induced non-locality to the standard optimization question governed by the
the inverse approximate joint diagonalization. In Sec. IV, we demonstrate the power of the new definition and the
algorithm by various applications. The conclusion is obtained in Sec. V.

\section{The measurement-induced non-locality based on the skew information}

\textit{The definition.-}To begin with, we briefly introduce the measurement-induced non-locality \cite%
{loca} based on the $l_{2}$ norm. For a bipartite density matrix $\rho _{AB}$%
, the measurement-induced non-locality is defined by the maximal disturbance of local projective
measurements as
\begin{equation}
N(\rho _{AB})=\max_{\Pi }\left\Vert \rho -\Pi ^{A}(\rho _{AB})\right\Vert
_{2}^{2},  \label{MIND}
\end{equation}%
where $\left\Vert \cdot \right\Vert _{2}$ denotes the $l_{2}$ norm of a
matrix, the von Neumann measurement $\Pi ^{A}$ is defined by $\sum_{k}(\Pi
_{k}^{A}\otimes \mathbb{I}^{B})\rho _{AB}(\Pi _{k}^{A}\otimes \mathbb{I}%
^{B}) $ with $\sum_{k}\Pi _{k}^{A}\rho _{A}\Pi _{k}^{A}=\rho _{A}$
guaranteeing that the reduced density matrix $\rho _{A}=\mathrm{Tr}_{B}\rho
_{AB}$ is not disturbed. The measurement-induced non-locality can be analytically calculated for
two-qubit systems, but it is not contractive in that the measurement-induced non-locality can be
increased by some local operations on subsystem B and decreased by the
tensor product of a third independent subsystem. In order to avoid the
non-contractivity, we can effectively utilize the properties of the skew
information and directly present our definition of the measurement-induced non-locality in the following
rigorous way.

\textbf{Definition 1.-} For an ($m\otimes n$)-dimensional density $\rho _{AB}$%
, the measurement-induced non-locality can be defined in terms of skew information as
\begin{equation}
U(\rho _{AB})=\max_{\left\{ K_{k}\right\}
}\sum_{k=0}^{m-1}I\left(\rho_{AB},K_k\right),  \label{U}
\end{equation}%
where $I\left(\rho_{AB},K_k\right)=-\frac{1}{2}\mathrm{Tr}[\sqrt{\rho _{AB}}%
,K_{k}]^{2}$ is the quantum skew information and $K_{k}=\left\vert
k\right\rangle _A\left\langle k\right\vert \otimes \mathbb{I}_B$ with $%
\left\vert k\right\rangle $ denoting the eigenvectors of the reduced density
matrix $\rho _{A}$.

At first, one can find that (i) $U(\rho_{AB})=0$ for any product state $\rho_{AB}=\rho_{A}\otimes\rho_{B}$;
(ii) $U(\rho _{AB})$  is invariant under local unitary transformations;
(iii) $\rho _{AB}$ vanishes for any classical-quantum state $\rho _{AB}=\sum_kp_k\left\vert k\right\rangle_A\left\langle k\right\vert\otimes\rho_B$ with the nondegenerate reduced density matrix $\rho_A$; (iv) $U(\rho _{AB})$ is equivalent to the entanglement for pure $\rho _{AB} $, which can be found from our latter Eq. (\ref{pure}). All the above properties of our redefined $U(\rho_{AB})$ are completely the same as the fundamental properties of the measurement-induced non-locality given in Ref. \cite{loca}. This means that our $U(\rho_{AB})$ characterizes the same quantum resource as that in Ref. \cite{loca}. In addition, it is obvious that $U(\rho _{AB})$ is contractive and it is invariant under
local unitary operations as a result of the good properties of the skew
information. In addition, it can be found that the above definition of measurement-induced non-locality
has the obvious operational meaning related to quantum metrology.

\textit{Operational meaning.-} Let's consider a scheme of quantum metrology
as follows. Suppose we have an $(m\otimes n)$-dimensional state $\rho_{AB}$
with the reduced density matrix $\rho_A$ and then let the state undergo a
unitary operation $U_{\varphi _{k}}=e^{-iK_k\varphi _{k}}$ with $%
K_{k}=\left\vert k\right\rangle _A\left\langle k\right\vert \otimes \mathbb{I%
}_B$ and $\left[\left\vert k\right\rangle\left\langle k\right\vert,\rho_A%
\right]=0$. This will endow an unknown phase $\varphi _{k}$ to the state $%
\rho_{AB} $ as $\rho _{k}=U_{\varphi _{k}}\rho_{AB} U_{\varphi
_{k}}^{\dagger }$. We aim to estimate the measurement precision by measuring
$\varphi _{k}$ in $\rho _{k}$ with $N>>1$ runs of detection on $\rho _{k}$.

In the above scheme, the measurement precision of $\varphi _{k}$ is
characterized by the uncertainty of the estimated phase $\varphi _{k}^{est}$
defined by
\begin{equation}
\delta \varphi _{k}=\left\langle \left( \frac{\varphi _{k}^{est}}{\left\vert
\partial \left\langle \varphi _{k}^{est}\right\rangle /\partial \varphi
_{k}\right\vert }-\varphi _{k}\right) ^{2}\right\rangle ^{1/2}
\end{equation}%
which, for an unbiased estimator, is just the standard deviation \cite%
{fis2,fis1,fis11}. Based on the quantum parameter estimation\cite%
{fis2,fis1,fis11}, $\delta \varphi _{k}$ is limited by the quantum Cram\'{e}%
r-Rao bound as $\left( \delta \varphi _{k}\right) ^{2}\geq \frac{1}{NF_{Qk}}$%
, where $F_{Qk}=\mathrm{Tr}\{\rho _{\varphi }L_{\varphi }^{2}\}$ is the
quantum Fisher information with $L_{\varphi }$ being the symmetric
logarithmic derivative defined by $2\partial _{\varphi }\rho _{\varphi
}=L_{\varphi }\rho _{\varphi }+\rho _{\varphi }L_{\varphi }$ \cite{fis2}. It
was shown in Refs. \cite{fis2,fis1,fis11} that this bound can always be
reached asymptotically by maximum likelihood estimation and a projective
measurement in the eigen-basis of the "symmetric logarithmic derivative
operator" . Thus one can let $\left( \delta \varphi _{k}^{o}\right) ^{2}$ to
denote the optimal variance which achieves the Cram\'{e}r-Rao bound, i.e., $%
\left( \delta \varphi _{k}^{o}\right) ^{2}=\frac{1}{NF_{Qk}}$. Ref. \cite%
{luo1} showed that the Fisher information $F_{Qk}$ is well bounded by the
skew information as
\begin{equation}
\frac{F_{Qk}}{4}\leq 2I\left( \rho_{AB} ,K_k \right)\iff \frac{1}{\left(
\delta \varphi _{k}^{o}\right) ^{2}}\leq 8NI(\rho_{AB},K_k ).  \label{resufi}
\end{equation}

Suppose we repeat this scheme $N$ times respectively corresponding to a
complete set $\{\left.K_k\right\vert\left[\left\vert k\right\rangle
\left\langle k\right\vert,\rho_A\right]=0 \}$, we can sum Eq. (\ref{resufi})
over $k$ as
\begin{equation}
\sum_{k}\frac{1}{\left( \delta \varphi _{k}^{o}\right) ^{2}}\leq
8N\sum_{k}I(\rho_{AB},K_k )\leq8NU(\rho_{AB}),  \label{fre}
\end{equation}%
where the last inequality comes from Eq. (\ref{U}). If we define $\frac{1}{%
(\Delta _{\varphi }^{o})^{2}}=\sum_{k}\frac{1}{\left( \delta \varphi
_{k}^{o}\right) ^{2}}$, Eq. (\ref{fre}) can be rewritten as
\begin{equation}
\frac{1}{8NU(\rho_{AB} )}\leq (\Delta _{\varphi }^{o})^{2 },  \label{general}
\end{equation}%
which shows that the measurement-induced non-locality $U(\rho_{AB} )$ contributes to the lower bounds of
the ``average variance" $(\Delta _{\varphi }^{o})^{2}$ that characterizes
the contributions of all the inverse optimal variances of the estimated
phases. It is worth emphasizing that $(\Delta _{\varphi }^{o})^{2}$ \textit{%
corresponds to an arbitrary complete set $\{\left.K_k\right\vert\left[%
\left\vert k\right\rangle \left\langle k\right\vert,\rho_A\right]=0 \}$
instead of the optimal set in the sense of Eq. (\ref{U})}. In addition,  in the above mentioned asymptotical and the optimal regimes, Eq. (\ref{general}) for the pure states will become $\frac{1}{4NU(\rho_{AB} )}= (\Delta _{\varphi }^{o})^{2 }$ due to the inequality $\frac{F_{Qk}}{4}\geq I\left( \rho_{AB}\right)$ \cite{luo1}. However, one
can recognize that in the practical scenario, $\delta\varphi_k\geq\delta%
\varphi^o_k$ because the measurement protocol couldn't be optimal. Thus, one
can immediately obtain
\begin{equation}
\sum_{k}\frac{1}{\left( \delta \varphi _{k}\right) ^{2}}\leq 8NU\left(
\rho_{AB}\right)\iff\frac{1}{8NU(\rho_{AB} )}\leq (\Delta _{\varphi })^{2 },
\label{general2}
\end{equation}
with $\frac{1}{(\Delta _{\varphi })^{2}}=\sum_{k}\frac{1}{\left( \delta
\varphi _{k}\right) ^{2}}$. Thus we would like to emphasize that Eq. (\ref{general2}) and Eq. (\ref{general})
provide only the lower bound instead of the exact value of the ``average variance" $(\Delta _{\varphi })^{2}$ or $(\Delta _{\varphi }^{o})^{2}$, because the inequalities are not saturated in the general cases. In this sense, our measurement-induced non-locality was endowed with the operational meaning.

\section{The effective numerical algorithm and the measurement-induced non-locality in arbitrary dimension%
}

\textit{The approximate joint diagonalization algorithm.-} In order to give our main inverse approximate joint diagonalization algorithm, we
first introduce the well-known approximate joint diagonalization algorithm \cite{blind,jacibo,fast} and
its relevant approximate joint diagonalization optimization problem. For $m$ \textit{n}-dimensional
matrices $\{M_{m}\}$, one aims to find a unitary matrix $U$ such that
\begin{equation}
J(U)=\max_{U}\sum_{i=1}^{m}\sum_{k=1}^{n}|UM_{i}U^{\dagger }|_{kk}^{2}.
\label{J(U)}
\end{equation}
Such an optimization problem is a standard expression of the approximate joint diagonalizationapproximate joint diagonalization of the series
of matrices $M_{i}$. This approximate joint diagonalization problem is widely met in the Blind Source
Separation and Independent Component Analysis (See Ref. \cite{golub} and the
references therein) and is well solved numerically by many effective
algorithms . A very remarkable algorithm is the Jocabi method \cite%
{blind,jacibo,golub} which can be performed as steadily, reliably, fast and
perfectly as the diagonalization of a single matrix. The main idea of the approximate joint diagonalization
algorithm of our interest is as follows.

Any unitary matrix $U$ can be decomposed into a series of unitary matrices
called Givens rotations as $U=\Pi _{\{\theta ,\phi \}}U(\theta ,\phi )$ with
\begin{equation}
U(\theta ,\phi )=%
\begin{pmatrix}
\cos \theta  & e^{i\varphi }\sin \theta  \\
-e^{-i\varphi }\sin \theta  & \cos \theta
\end{pmatrix}%
.
\end{equation}%
Each Givens rotation is only operated on the $(2\times 2)$-dimensional
subspace of the $m$ $M_{i}$'s. Let
\begin{equation}
G_{i}=%
\begin{pmatrix}
a_{i} & b_{i} \\
c_{i} & d_{i}%
\end{pmatrix}%
\end{equation}%
denote the $(2\times 2)$-dimensional block matrix of $M_{i}$ corresponding
to the Givens rotation $U(\theta ,\phi )$. After the Givens rotation, $G_{i}$
is updated by
\begin{equation}
G_{i}^{\prime }=U^{\dagger }(\theta ,\phi )G_{i}U(\theta ,\phi )=%
\begin{pmatrix}
a_{i}^{\prime } & b_{i}^{\prime } \\
c_{i}^{\prime } & d_{i}^{\prime }%
\end{pmatrix}%
.
\end{equation}%
In this $(2\times 2)$-dimensional subspace, the optimization problem Eq. (%
\ref{J(U)}) is equivalent to the problem
\begin{eqnarray}
&&\max \left\{ \left\vert a_{i}^{\prime }\right\vert ^{2}+\left\vert
d_{i}^{\prime }\right\vert ^{2}\right\}  \\
&=&\max \left\{ \frac{1}{2}\left\vert a_{k}^{\prime }-d_{k}^{\prime
}\right\vert ^{2}+\frac{1}{2}\left\vert a_{k}^{\prime }+d_{k}^{\prime
}\right\vert ^{2}\right\} ,  \notag  \label{op2}
\end{eqnarray}%
by choosing suitable parameter $\theta $ and $\varphi $. Since the trace of
a matrix is preserved under the unitary operations, i.e., $a_{k}^{\prime
}+d_{k}^{\prime }=a_{k}+d_{k}$, the optimization problem Eq. (\ref{op2}) is
equivalent to
\begin{equation}
Q=\max \sum_{i=1}^{m}\left\vert a_{k}^{\prime }-d_{k}^{\prime }\right\vert
^{2},
\end{equation}%
which can be rewritten as \cite{jacibo}
\begin{equation}
Q=\max \frac{v^{\intercal }\mathrm{Re}(G^{\dagger }G)v}{v^{\intercal }v},
\label{qqqq}
\end{equation}%
where $G=[g_{1},g_{2},...,g_{m}]$, $%
g_{k}=[a_{k}-d_{k},b_{k}+c_{k},i(c_{k}-b_{k})]^{\intercal }$, $v=[\cos
2\theta ,-\sin 2\theta \cos \varphi ,-\sin 2\theta \sin \varphi ]^{\intercal
}$. The optimization problem maximizing $Q$ in Eq. (\ref{qqqq}) happens to
be the well-known Rayleigh Quotient which shows that $Q$ has a closed form
and is just the maximal eigenvalue of the matrix $\mathrm{Re}(G^{\dagger }G)$%
. The eigenvector for the maximal eigenvalue achieves the optimization
solution, by which one can solve the corresponding $(\theta ,\phi )$ and
further find the optimal Givens rotation $U(\theta ,\phi )$.

When all $(2\times 2)$-dimensional block matrices in $M_i$ are updated once
by the Givens rotations, it is called as one sweep. Following the same
procedure, each sweep transfers the contribution of the off diagonal entries
to the diagonal entries as required by optimization problem Eq. (\ref{J(U)}%
). The sweep is performed again and again until the required precision is
reached.

\textit{The inverse approximate joint diagonalization algorithm.-} The inverse approximate joint diagonalization algorithm is completely parallel with
the approximate joint diagonalization algorithm. Let's consider an inverse problem opposite to the problem Eq.
(\ref{J(U)}), namely, finding a proper unitary operation $U$ such that
\begin{equation}
\tilde{J}(U)=\min_{U}\sum_{i=1}^{m}\sum_{k=1}^{n}|UM_{i}U^{\dagger
}|_{kk}^{2}.  \label{J(u)bar}
\end{equation}%
Following the completely same procedure as the approximate joint diagonalization algorithm, we can finally
arrive at
\begin{equation}
\tilde{Q}=\min \frac{v^{\intercal }\mathrm{Re}(G^{\dagger }G)v}{v^{\intercal
}v}.  \label{qqqqq}
\end{equation}%
Based on the Rayleigh Quotient, $\tilde{Q}$, dual to $Q$, is just the
minimal eigenvalue of the matrix $\mathrm{Re}(G^{\dagger }G)$. That is, $%
\tilde{J}(U)$ can be completely solved if replacing the maximal eigenvalue
in the approximate joint diagonalization algorithm by the minimal eigenvalue. This is the inverse approximate joint diagonalization algorithm.

Based on the above algorithms, one can always find the optimal unitary
operation $U$ subject to a satisfactory precision. Corresponding, one can
calculate the exact value of $\tilde{J}(U)$ or ${J}(U)$. For a neat
formulation, we would like to give another definition.

\textbf{Definition 2.-} The \textit{inverse joint diagonalizer} of \textit{m}
matrices $M_{i}$ is defined by $U_{o}$ which is the optimal unitary
operation that achieves the aim of Eq. (\ref{J(u)bar}), i.e., $\tilde{J}%
(U)=\min_{U_{0}}\sum_{i=1}^{m}\sum_{k=1}^{n}\left\vert
U_{o}M_{i}U_{o}^{\dagger }\right\vert _{kk}^{2}$. The \textit{inverse joint
eigenvalue} of $M_{i}$ is defined by $\tilde{\lambda}_{k}^{i}=\left[
U_{o}M_{i}U_{o}^{\dagger }\right] _{kk}$.

\textit{The measurement-induced non-locality for arbitrary quantum states.-} Based on Definition 2, we can
present our main theorem as follows.

\textbf{Theorem 1.-} For an $(m\otimes n)$-dimensional state $\rho _{AB}$,
define the matrices $A_{ij}=(\mathbb{I}_{m}\otimes \left\langle
\psi_i\right\vert )\sqrt{\rho _{AB}}(\mathbb{I}_{m}\otimes \left\vert
\psi_j\right\rangle )$ with $\left\{\left\vert \psi_i\right
\rangle\right\}$
denoting any orthonormal basis of subsystem $B$. Then the measurement-induced non-locality $U(\rho _{AB})$
of $\rho _{AB}$ can be given by
\begin{equation}
U(\rho _{AB})=1-\sum_{i,j=0}^{n-1}\left\{\sum_{k=1}^N\left\vert
\Phi^\dagger_NA_{ij}\Phi_N\right\vert^{2}_{kk}+\sum_{\alpha}\sum_{k=1}^{D_%
\alpha}\left\vert\tilde{\lambda} _{k}^{\alpha ij}\right\vert^{2}\right\},
\label{theorem}
\end{equation}%
where $\Phi _{N}=[\left\vert 1\right\rangle ,\left\vert
2\right\rangle,\cdots,\left\vert N\right\rangle ]$ is the matrix made up of
all the non-degenerate eigenvectors $\left\{\left\vert
k\right\rangle\right\} $ of the reduced density matrix $\rho_A=\mathrm{Tr}%
_B\rho_{AB}$, $\Phi _{D_\alpha}=[\left\vert 1\right\rangle
_\alpha,\left\vert 2\right\rangle _\alpha,\cdots,\left\vert D\right\rangle
_\alpha ]$ is the matrix made up of all the eigenvectors in the $\alpha$th
degenerate subspace of $\rho_A$ with $N+\sum_\alpha D_\alpha=m$ and $\tilde{%
\lambda} _{k}^{\alpha ij}$ is the $k$th inverse joint eigenvalue of the
matrix $\Phi_{D_\alpha}^\dagger A_{ij}\Phi_{D_\alpha}$.

\textbf{Proof.} From Eq. (\ref{U}), we can find
\begin{eqnarray}
U(\rho _{AB}) &=&\max_{K_{k}}\sum_{k=0}^{m-1}[\mathrm{Tr}\rho _{AB}K_{k}^{2}-%
\mathrm{Tr}\sqrt{\rho _{AB}}K_{k}\sqrt{\rho _{AB}}K_{k}]  \notag \\
&=&1-\min_{K_{k}}\sum_{k=0}^{m-1}\mathrm{Tr}\sqrt{\rho _{AB}}K_{k}\sqrt{\rho
_{AB}}K_{k}.  \label{one}
\end{eqnarray}%
Substituting $K_{k}=\left\vert k\right\rangle \left\langle k\right\vert
\otimes \mathbb{I}$ with $\left\{ \left\vert k\right\rangle \right\} $
denoting the eigenvectors of the reduced density matrix $\rho _{A}$ and any
orthonormal bases $\left\{ \left\vert \psi _{i}\right\rangle \right\} $ of
subsystem B into Eq. (\ref{one}), it follows that%
\begin{eqnarray}
U(\rho _{AB}) &=&1-\min_{\{\left\vert k\right\rangle
\}}\sum_{i,j=0}^{n-1}\sum_{k=0}^{m-1}\mathrm{Tr}\sqrt{\rho _{AB}}(\left\vert
k\right\rangle \left\langle k\right\vert \otimes \left\vert \psi
_{i}\right\rangle \left\langle \psi _{i}\right\vert )  \notag \\
&&\times \sqrt{\rho _{AB}}(\left\vert k\right\rangle \left\langle
k\right\vert \otimes \left\vert \psi _{j}\right\rangle \left\langle \psi
_{j}\right\vert )  \notag \\
&=&1-\min_{\{\left\vert k\right\rangle
\}}\sum_{i,j=0}^{n-1}\sum_{k=0}^{m-1}|\left\langle k\right\vert
A_{ij}\left\vert k\right\rangle |^{2},  \label{firuu}
\end{eqnarray}%
with $A_{ij}=(\mathbb{I}_{m}\otimes \left\langle \psi _{i}\right\vert )\sqrt{%
\rho _{AB}}(\mathbb{I}_{m}\otimes \left\vert \psi _{j}\right\rangle )$.
Suppose that the reduced density matrix $\rho _{A}$ has $N$ non-degenerate
eigenvalues and $\alpha $ degenerate subspace with the dimension denoted by $%
D_{\alpha }$, respectively, then one can always construct an $(m\times N)$%
-dimensional matrix $\Phi _{N}$ as $\Phi _{N}=[\left\vert 1\right\rangle
,\left\vert 2\right\rangle ,\cdots ,\left\vert N\right\rangle ]$ with the
column $\left\vert k\right\rangle $ including all the non-degenerate
eigenvectors of $\rho _{A}$ and $\alpha $ $(m\times D)$-dimensional matrices
$\Phi _{D_{\alpha }}$ as $\Phi _{D_{\alpha }}=[\left\vert 1\right\rangle
_{\alpha },\left\vert 2\right\rangle _{\alpha },\cdots ,\left\vert
D\right\rangle _{\alpha }]$ with the column $\left\vert k\right\rangle
_{\alpha }$ covering all the eigenvectors in the $\alpha $th degenerate
subspace of $\rho _{A}$. It is obvious that $N+\sum_{\alpha }D_{\alpha }=m$.
Thus the optimization of $\{\left\vert k\right\rangle \}$ in Eq. (\ref{firuu}%
) is converted to the optimization of $\Phi _{N}$ and $\Phi _{D_{\alpha }}$.
However, $\Phi _{N}$ is made up of the non-degenerate eigenvectors of $\rho
_{A}$, so $\Phi _{N}$ is uniquely determined by $\rho _{A}$. As a result,
only the optimization of $\Phi _{D_{\alpha }}$ is required, since $\Phi
_{D_{\alpha }}$ is not unique in the degenerate subspace. If we denote any
two different such matrices as $\Phi _{D_{\alpha }}$ and $\Psi _{D_{\alpha
}} $, there always exists a unitary transformation as $\Psi _{D_{\alpha
}}=\Phi _{D_{\alpha }}U$. So the optimization of $\Phi _{D_{\alpha }}$ is
essentially converted to the optimization of the unitary operations $U$ once
one fixes a representation (a particular $\Phi _{D_{\alpha }}$ ). Thus Eq. (%
\ref{firuu}) can be rewritten as
\begin{equation}
U(\rho _{AB})=1-\sum_{i,j=0}^{n-1}\sum_{k=1}^{N}\left\vert \Phi
_{N}^{\dagger }A_{ij}\Phi _{N}\right\vert _{kk}^{2}-\sum_{\alpha }\tilde{J}%
_{\alpha }(U_{\alpha }),  \label{lastok}
\end{equation}%
with
\begin{equation}
\tilde{J}_{\alpha }(U_{\alpha })=\min_{U_{\alpha
}}\sum_{i,j=0}^{n-1}\sum_{k=1}^{D_{\alpha }}\left\vert U_{\alpha }^{\dagger
}\Phi _{D_{\alpha }}^{\dagger }A_{ij}\Phi _{D_{\alpha }}U_{\alpha
}\right\vert _{kk}^{2}.  \label{opt}
\end{equation}%
It is obvious that Eq. (\ref{opt}) is the standard inverse approximate joint diagonalization problem of matrices $%
\Phi _{D_{\alpha }}^{\dagger }A_{ij}\Phi _{D_{\alpha }}$ given in Eq. (\ref%
{J(u)bar}) which can be well solved by the inverse approximate joint diagonalization algorithm. Let $\lambda
_{k}^{\alpha ij}$ denote the $k$th inverse joint eigenvalue of the matrix $%
\Phi _{D_{\alpha }}^{\dagger }A_{ij}\Phi _{D_{\alpha }}$, then Eq. (\ref%
{lastok}) exactly becomes Eq. (\ref{theorem}) in our main theorem, which
completes our proof. \hfill \rule{2.5mm}{2.5mm}

\textit{Efficiency of the inverse approximate joint diagonalization algorithm and the closure of the measurement-induced non-locality.-} From the
approximate joint diagonalization and inverse approximate joint diagonalization algorithms in the above paragraph, one can find that the two
algorithms have the equal efficiency within the same precision. It is
especially noted that the approximate joint diagonalization algorithm for a single matrix is exactly the
Jacobi method for the diagonalization of a single matrix \cite{vanloanbook}.
In this sense, we can safely draw the conclusion that the inverse approximate joint diagonalization algorithm has
the same efficiency as the Jacobi method if they are executed for the same
dimensional matrix subject to the same precision. Now let's consider the
inverse approximate joint diagonalization algorithm in our the measurement-induced non-locality for an $(m\otimes n)$-dimensional density matrix $%
\rho _{AB}$. It is equivalent to the inverse approximate joint diagonalization of $n^{2}$ $(m\times m)$-dimensional
matrices $A_{ij}$ in the worst case (the reduced matrix is a maximally mixed
state). Thus one can find that one sweep needs $\frac{m(m-1)}{2}$ Given
rotations and $\frac{m^{2}n^{2}-mn^{2}}{2}$ unitary transformations (Givens
rotation operations). However, if one diagonalizes $\rho _{AB}$ with the
Jacobi method, one will require $\frac{mn(mn-1)}{2}$ Given rotations and the
equal number of unitary transformations. It is obvious that under the same
condition, the inverse approximate joint diagonalization algorithm requires much less Givens rotations and
unitary transformations than the Jacobi method for the diagonalization of $%
\rho _{AB}$ especially for large $n$, which shows the high efficiency of the
IADJ algorithm. In particular, it was said that about $\log (m)$ sweeps were
needed for a single matrix $M_{i}$ subject to an acceptable precision
(Examples show around $10^{-16}$ in \cite{vanloanbook}).

In addition, one always expects a closed form (or an analytic expression) of
a quantifier. This could be, in principle, achieved if the quantifier could
be given in terms of the eigenvalues of some matrices, since the eigenvalues
can be completely determined by the secular equation of the matrix no matter
whether they can be exactly calculated or not. However, in the practical
scenario, the exact eigenvalues of a matrix can never be obtained especially
for the high dimensional case. Thus a numerical algorithm is inevitable,
which will have to lead to the practical broken closure. In other words, in
the practical case, our inverse joint eigenvalues used above have the equal
weight as the eigenvalues of some matrices. In this sense, we emphasize that
our expression in theorem 1 is almost analytic (or of closed form).

\section{Examples and the power of the inverse approximate joint diagonalization algorithm}

Next, we would like to give some examples to verify the effectiveness of our
theorem 1 by comparing the strictly analytic expression and the numerical
ones in theorem 1. All the examples show the perfect consistency and prove
the efficiency and the superiority of our theorem 1 especially in the high
dimensional case.

\textit{(i) Bipartite pure states.-} Any a bipartite pure state can be given
in the Schmidt decomposition as $\left\vert \chi \right\rangle
_{ab}=\sum_{i}u_{i}\left\vert ii\right\rangle _{ab}$ where $u_{i}$ is the
Schmidt coefficients. The measurement-induced non-locality can be easily obtained as
\begin{eqnarray}
U(\rho _{ab}) &=&1-\min_{K_{k}}\sum_{k=0}^{n-1}Tr(\sqrt{\rho _{AB}}K_{k}%
\sqrt{\rho _{AB}}K_{k})  \notag \\
&=&1-\min_{\{\left\vert k\right\rangle \}}\sum_{k=0}^{n-1}\left\vert
\sum_{ij}u_{i}u_{j}\left\langle ii\right\vert (\left\vert k\right\rangle
\left\langle k\right\vert \otimes \mathbb{I}_{B})\left\vert jj\right\rangle
_{AB}\right\vert ^{2}  \notag \\
&=&1-\min_{\{\left\vert k\right\rangle \}}\sum_{k=0}^{n-1}\left\vert
\sum_{i}u_{i}^{2}\left\langle k\right. \left\vert i\right\rangle
\left\langle i\right. \left\vert k\right\rangle _{A}\right\vert ^{2}  \notag
\\
&=&1-\sum_{k=0}^{n-1}u_{k}^{4}=1-\mathrm{Tr}\rho _{A}^{2},  \label{pure}
\end{eqnarray}%
with the optimum value attained by $\left\langle k\right\vert \left\vert
i\right\rangle =\frac{1}{\sqrt{n}}$. It is explicit that the measurement-induced non-locality for a pure
state is exactly the half of its entanglement in terms of the linear entropy
of the reduced density matrix. In order to further validate our numerical
procedure in theorem 1, we plot Eq. (\ref{pure}) and its corresponding
numerical results for many pure states randomly generated by Matlab R2017a
in FIG. 1(a) which shows our numerical results are completely consistent
with Eq. (\ref{pure}).

\textit{(ii) Qubit-qudit states.-} For a $(2\otimes d)$-dimensional quantum
state $\rho _{AB}$ with the reduced density matrix $\rho _{A}$, the measurement-induced non-locality can
be given by
\begin{eqnarray}
U(\rho _{AB}) &=&-\frac{1}{2}\max_{[K_{k},\rho _{A}]=0}\sum_{k=0}^{1}\mathrm{%
Tr}[\sqrt{\rho _{AB}},K_{k}]^{2}  \notag \\
&=&-\max_{[K_{k},\rho _{A}]=0}\mathrm{Tr}[\sqrt{\rho _{AB}},K_{0}]^{2}.
\label{QUBIT}
\end{eqnarray}%
In the qubit subsystem, the observable $K_{0}$ can be expanded in the Bloch
representation as $K_{0}=\frac{1}{2}(\mathbb{I}_{2}+\vec{n}\cdot \vec{\sigma}%
)\otimes \mathbb{I}_{d}$ and $\rho _{A}$ can be expanded as $\rho _{A}=\frac{%
1}{2}(\mathbb{I}_{2}+\vec{r}\cdot \vec{\sigma})\otimes \mathbb{I}_{d}$,
where $\vec{n}$ with $\left\Vert \vec{n}\right\Vert _{2}=1$ and $\vec{r}$
with $\left\Vert \vec{r}\right\Vert _{2}\leq 1$ denote the Bloch vector.
Thus the optimization condition $[K_{k},\rho _{A}]=0$ is equivalent to $\vec{%
r}\times \vec{n}=0$ and Eq. (\ref{QUBIT}) arrives at
\begin{eqnarray}
U(\rho _{AB}) &=&\frac{1}{2}-\frac{1}{2}\min_{\vec{n}}\sum_{ij}\mathrm{Tr}%
n_{i}T_{ij}n_{j}  \notag \\
&=&\left\{
\begin{array}{cc}
\frac{1}{2}(1-v_{\min }), & \vec{r}=0 \\
\frac{1}{2}(1-\frac{1}{\left\Vert \vec{r}\right\Vert _{2}^{2}}\vec{r}^{T}T%
\vec{r}), & \vec{r}\neq 0%
\end{array}%
\right. ,  \label{2d}
\end{eqnarray}%
where $v_{\min }$ is the minimal eigenvalue of the matrix $T$ with $T_{ij}=%
\mathrm{Tr}\sqrt{\rho _{AB}}(\sigma _{i}\otimes \mathbb{I}_{n})\sqrt{\rho
_{AB}}(\sigma _{j}\otimes \mathbb{I}_{n})$. Eq. (\ref{2d}) gives the closed
form of the measurement-induced non-locality. Similar to the case (i), we also plot the numerical measurement-induced non-locality for
the randomly generated qubit-qudit states in FIG. 1(b) which also validates
our numerical results of theorem 1.
\begin{figure}[tbp]
\centering
\subfigure{\includegraphics[width=0.48\columnwidth,height=2in]{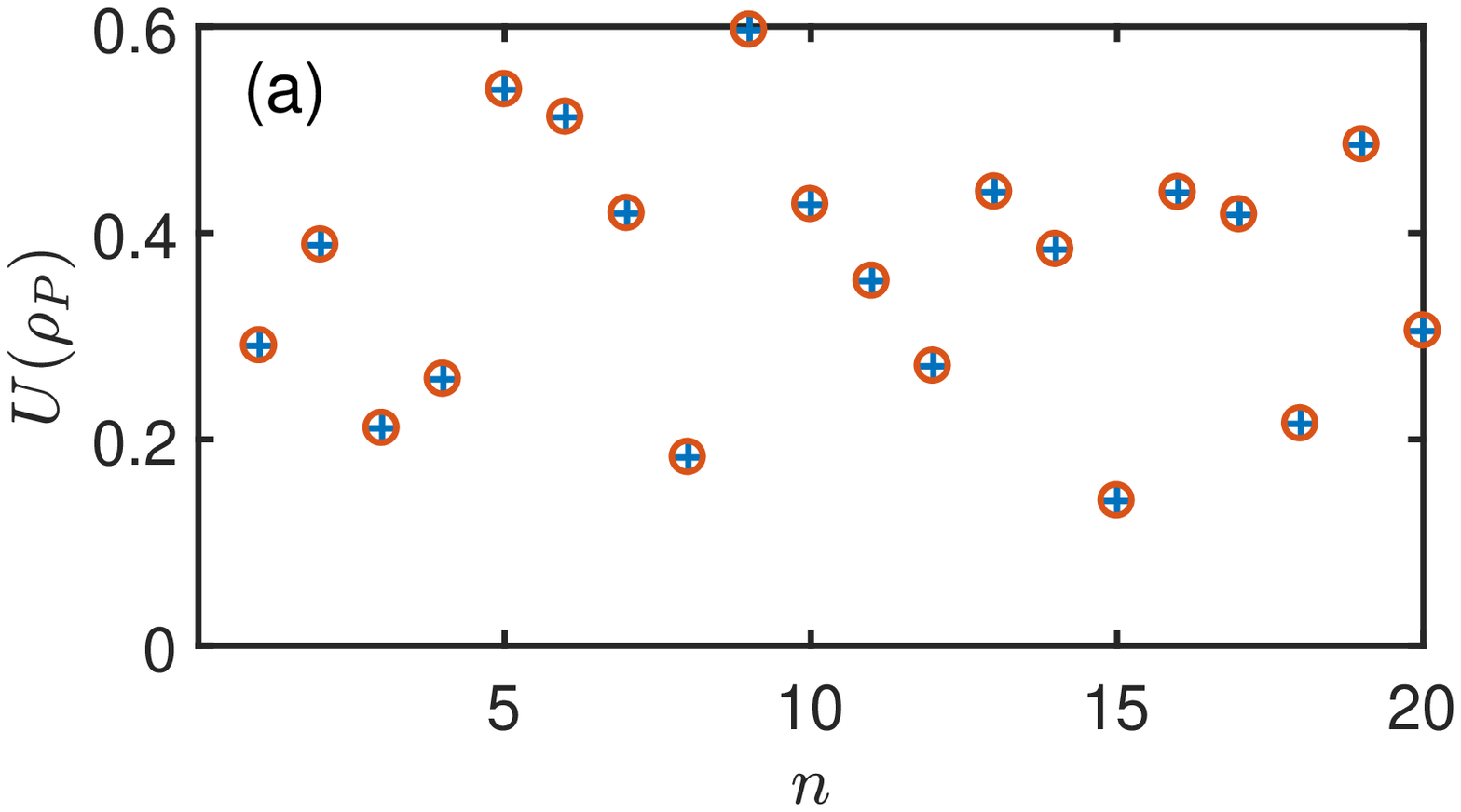}} %
\subfigure{\includegraphics[width=0.48\columnwidth,height=2in]{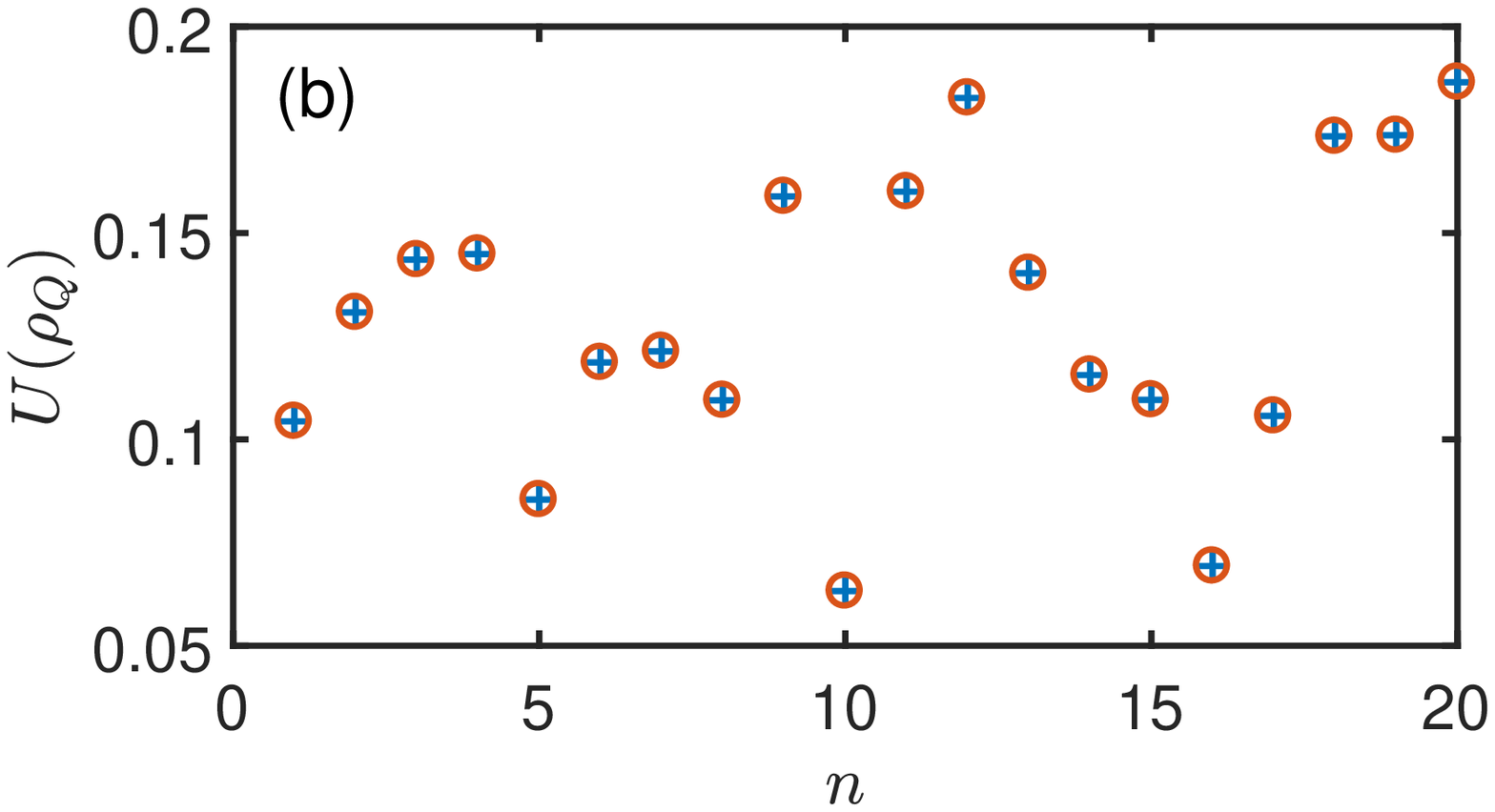}}%
\newline
\caption{(color online) The measurement-induced non-localities vs $n$. We, respectively, calculated 20 $%
(3\otimes3)$-dimensional pure states (a) and 20 $(2\otimes4)$-dime1nsional
qubit-qudit mixed states (b) randomly generated by Matlab R2014b. The "O"
stands for the strictly analytical expressions given by Eq. (\protect\ref%
{pure}) for (a) and Eq. {\protect\ref{2d}} for (b) and the "+" marks the
numerical expressions given by Eq. (\protect\ref{theorem}) in both figures.
Both show the perfect consistency. }
\end{figure}

\textit{(iii) The $(3\otimes 3)$-dimensional PPT states.-} The $(3\otimes 3)$%
-dimensional PPT states \cite{Alber} can be given by%
\begin{equation}
\rho _{PPT}=\frac{2}{7}\left\vert \Phi \right\rangle _{3}\left\langle \Phi
\right\vert +\frac{\alpha }{7}\rho _{+}+\frac{5-\alpha }{7}\rho _{-},\alpha
\in \lbrack 2,4],
\end{equation}%
where $\left\vert \Phi \right\rangle _{m}=\frac{1}{\sqrt{m}}%
\sum_{k=0}^{m-1}\left\vert kk\right\rangle $, and $\rho _{+}=\frac{1}{3}%
\sum_{k=0}^{2}\left\vert k,k\oplus 1\right\rangle \left\langle k,k\oplus
1\right\vert $and $\rho _{-}=\frac{1}{3}\sum_{k=0}^{2}\left\vert k\oplus
1,k\right\rangle \left\langle k\oplus 1,k\right\vert $ with $\oplus $ the
modulo-3 addition. The parameter $\alpha $ determines the different quantum
correlations of the PPT state. If $\alpha \leq 3$, $\rho _{ppt}$ is
separable. When $\alpha \in \lbrack 3,4]$, the PPT state is entangled. But
if $4\leq \alpha \leq 5$, $\rho _{PPT}$ is not a PPT state, but a free
entangled state. Based on our definition 1, one can find that $U(\rho _{AB})$
can be analytically calculated for $\alpha \in \lbrack 2,5]$ as
\begin{equation}
U(\rho _{PPT})=\left\{
\begin{array}{cc}
\frac{21-\sqrt{6(5-\alpha )}-\sqrt{6\alpha }-3\sqrt{\alpha (5-\alpha )}}{31.5%
}, & N_{T}<\alpha \leq 5 \\
\frac{4}{21}, & 2\leq \alpha \leq N_{T}%
\end{array}%
\right. ,  \label{ppt}
\end{equation}%
with $N_{T}=\frac{5+\sqrt{25-4\times (383-34\sqrt{94})/9}}{2}\approx 3.0669$%
. Both the numerical result based on our theorem and the strictly analytic
expression in Eq. (\ref{ppt}) are plotted in FIG. 2(a) which shows the
perfect consistency between them. In addition, we can analytically find the
sudden change point of the measurement-induced non-locality within the entanglement region. As a comparison,
we also plot the local quantum uncertainty (LQU) $Q(\rho _{PPT})$ given in
Ref. \cite{diagon} and the measurement-induced non-locality $U(\rho _{PPT})$ in FIG. 2(b). One can find that $%
U(\rho _{PPT}=Q(\rho _{PPT})$ at their common sudden change point and $%
U(\rho _{PPT}>Q(\rho _{PPT})$ as expected for other values of $\alpha $.
\begin{figure}[tbp]
\centering
\subfigure{\includegraphics[width=0.48\columnwidth,height=2in]{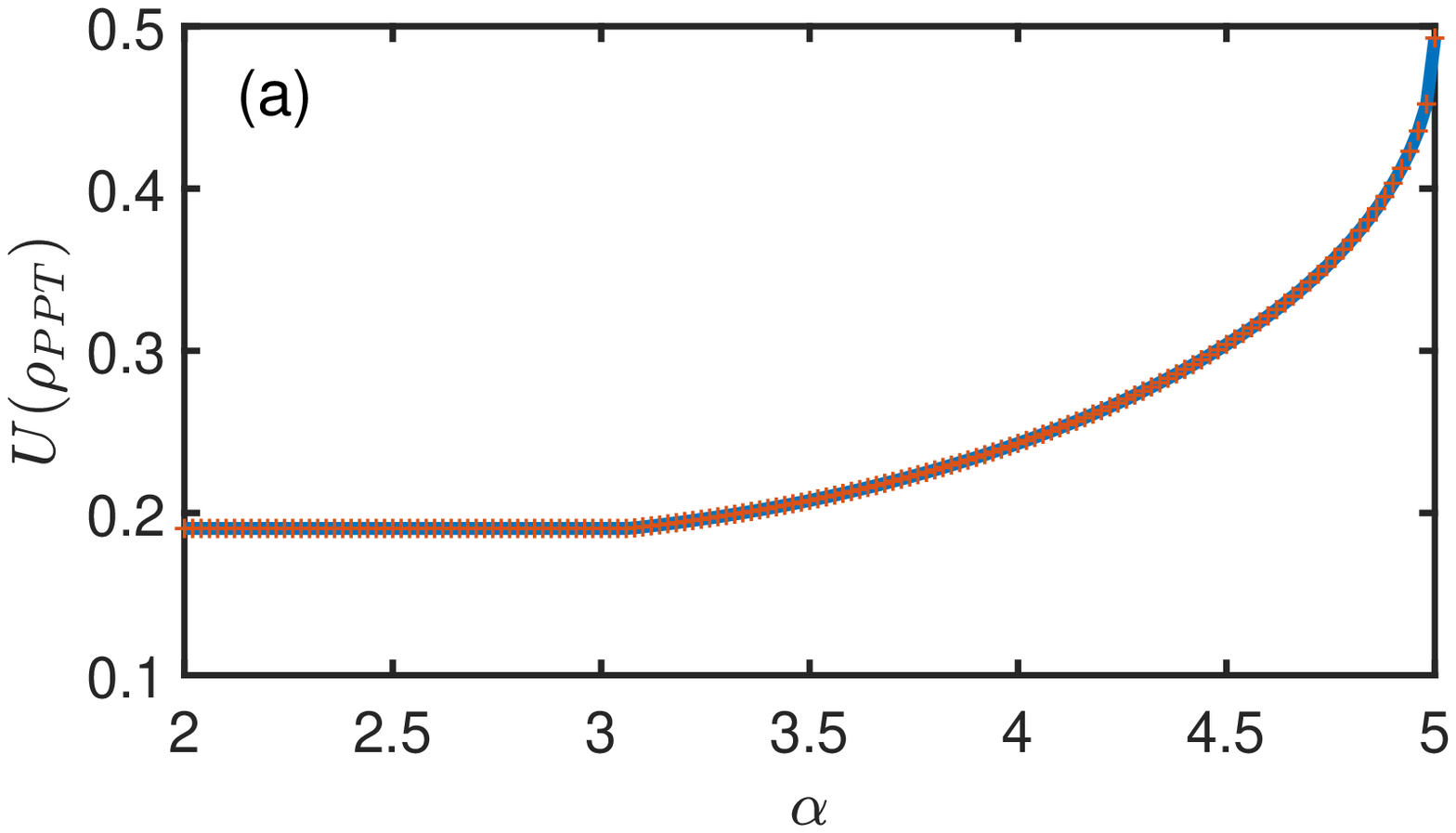}} %
\subfigure{\includegraphics[width=0.48\columnwidth,height=2in]{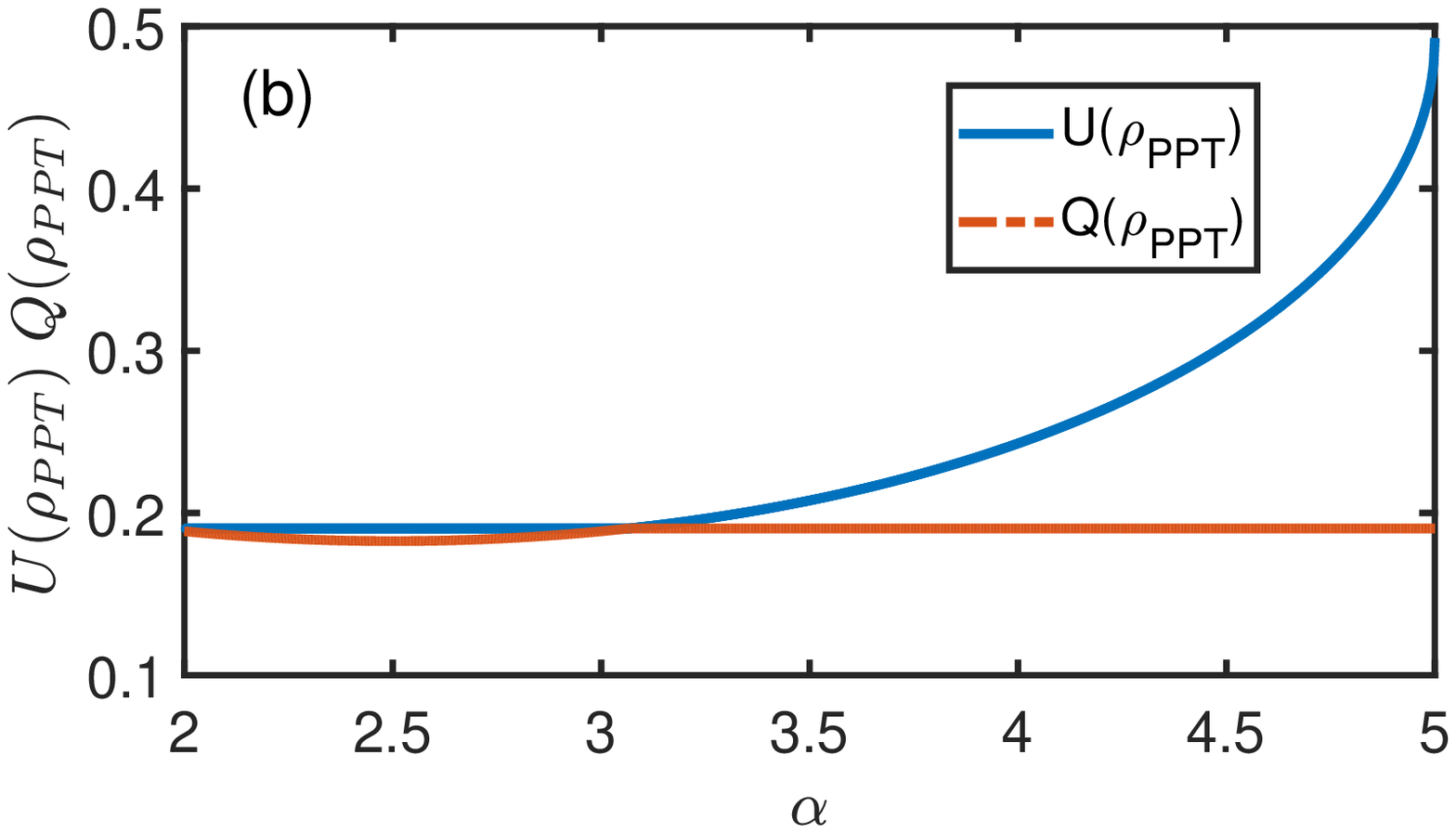}}%
\newline
\caption{(color online) The measurement-induced non-locality $U(\protect\rho _{PPT})$ for PPT states vs $%
\protect\alpha $ in (a). The solid line and the "+" line correspond to the
strictly analytical expression of Eq. (\protect\ref{ppt}) and the numerical
expression of Eq. (\protect\ref{theorem}). The sudden change point of the measurement-induced non-locality is
about $\protect\alpha \approx3.0669$. The measurement-induced non-locality is invariant for $\protect%
\alpha \leq 3.0669$, but becomes increasing for $\protect\alpha >3.0669$.
(b) shows the comparison of the LQU $Q(\protect\rho _{PPT})$ and the measurement-induced non-locality $U(%
\protect\rho _{PPT})$ based on PPT states. The solid line stands for $U(%
\protect\rho _{PPT})$, while the "-." line corresponds to $Q(\protect\rho %
_{PPT})$. These two lines share a common point at $\protect\alpha \approx
3.0669$. Except this point, $U(\protect\rho _{PPT})$ is always larger than $%
Q(\protect\rho _{PPT})$ as we expect.}
\end{figure}

\textit{(iv) The $(m\otimes m)$-dimensional isotropic states.-} The
isotropic states can be given by \cite{Alber}
\begin{equation}
\rho _{I}=\frac{1-x}{m^{2}-1}\mathbb{I}_{m^{2}}+\frac{m^{2}x-1}{m^{2}-1}%
\left\vert \Phi \right\rangle \left\langle \Phi \right\vert ,x\in \lbrack
-1,1],
\end{equation}%
with $\left\vert \Phi \right\rangle _{m}=\frac{1}{\sqrt{m}}%
\sum_{k=0}^{m-1}\left\vert kk\right\rangle $. Based on our definition, we
can analytically obtain%
\begin{equation}
U(\rho _{I})=\frac{m^{2}x-2x+1-2\sqrt{x(1-x)(m^{2}-1)}}{m(1+m)}.
\label{isotropic}
\end{equation}%
It is obvious that $U(\rho _{I})=0$ for $x=\frac{1}{m^{2}}$. As a
comparison, we plot $U(\rho _{I})$ given by Eq. (\ref{theorem}) and Eq. (\ref%
{isotropic}) in FIG. 3(a). The validity of our theorem is shown again.

\textit{(v) The $(m\otimes m)$-dimensional Werner states.-} The Werner
states can be written as \cite{Alber}
\begin{equation}
\rho _{W}=\frac{m-x}{m^{3}-m}\mathbb{I}_{m^{2}}+\frac{mx-1}{m^{3}-m}V,x\in
\lbrack -1,1],
\end{equation}%
with $V=\sum_{kl}\left\vert kl\right\rangle \left\langle lk\right\vert $ the
swap operator. The measurement-induced non-locality $U(\rho _{W})$ in terms of our definition can given
by
\begin{equation}
U(\rho _{W})=\frac{m-x-\sqrt{(m^{2}-1)(1-x^{2})}}{2(1+m)}.  \label{Werner}
\end{equation}%
From Eq. (\ref{Werner}), it is shown that $U(\rho _{W})=0$ for $x=\frac{1}{m}
$. The comparison between the numerical and the analytic expressions are
given in FIG. 3(b) which again shows the perfect consistency.
\begin{figure}[tbp]
\centering
\subfigure{\includegraphics[width=0.48\columnwidth,height=2in]{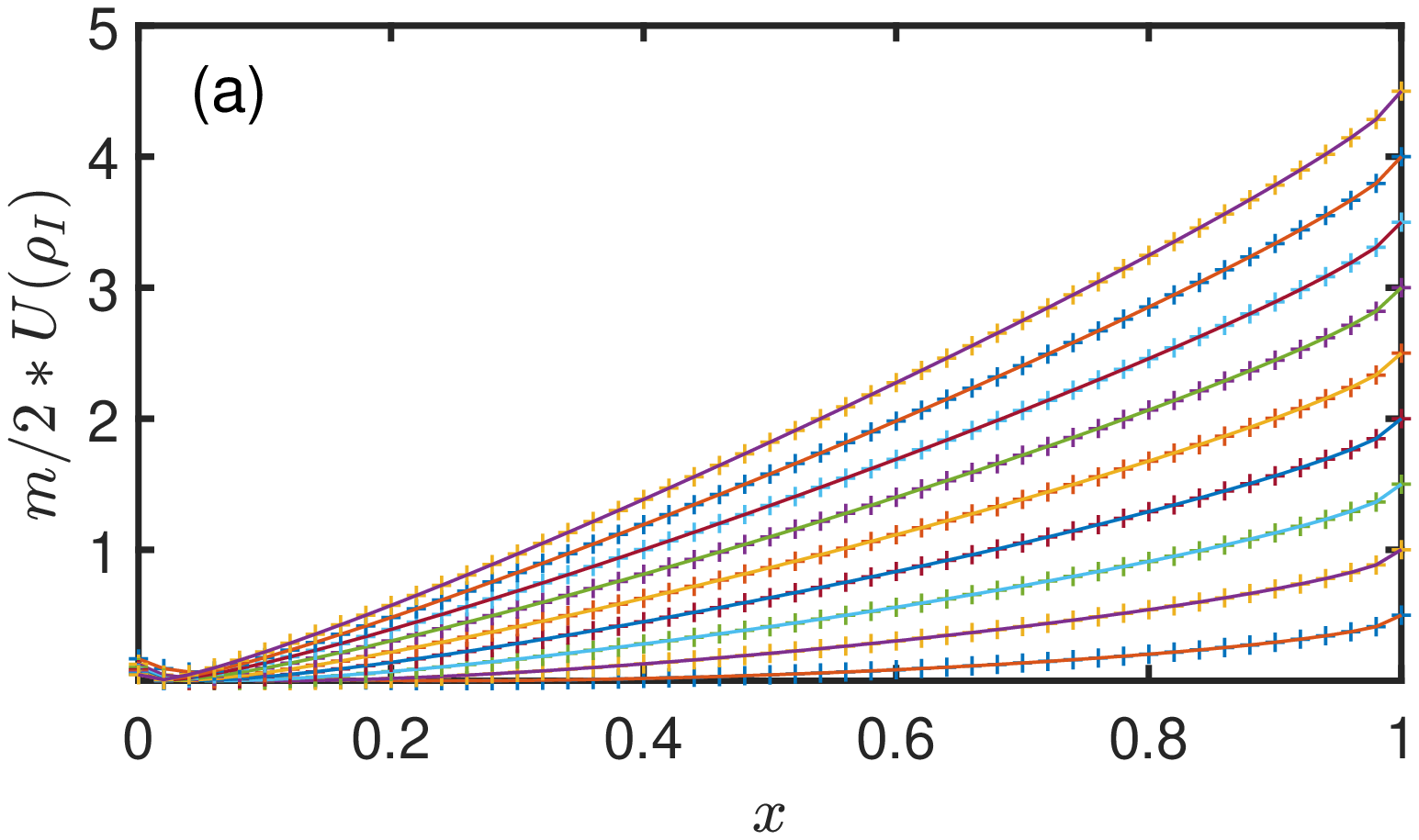}} %
\subfigure{\includegraphics[width=0.48\columnwidth,height=2in]{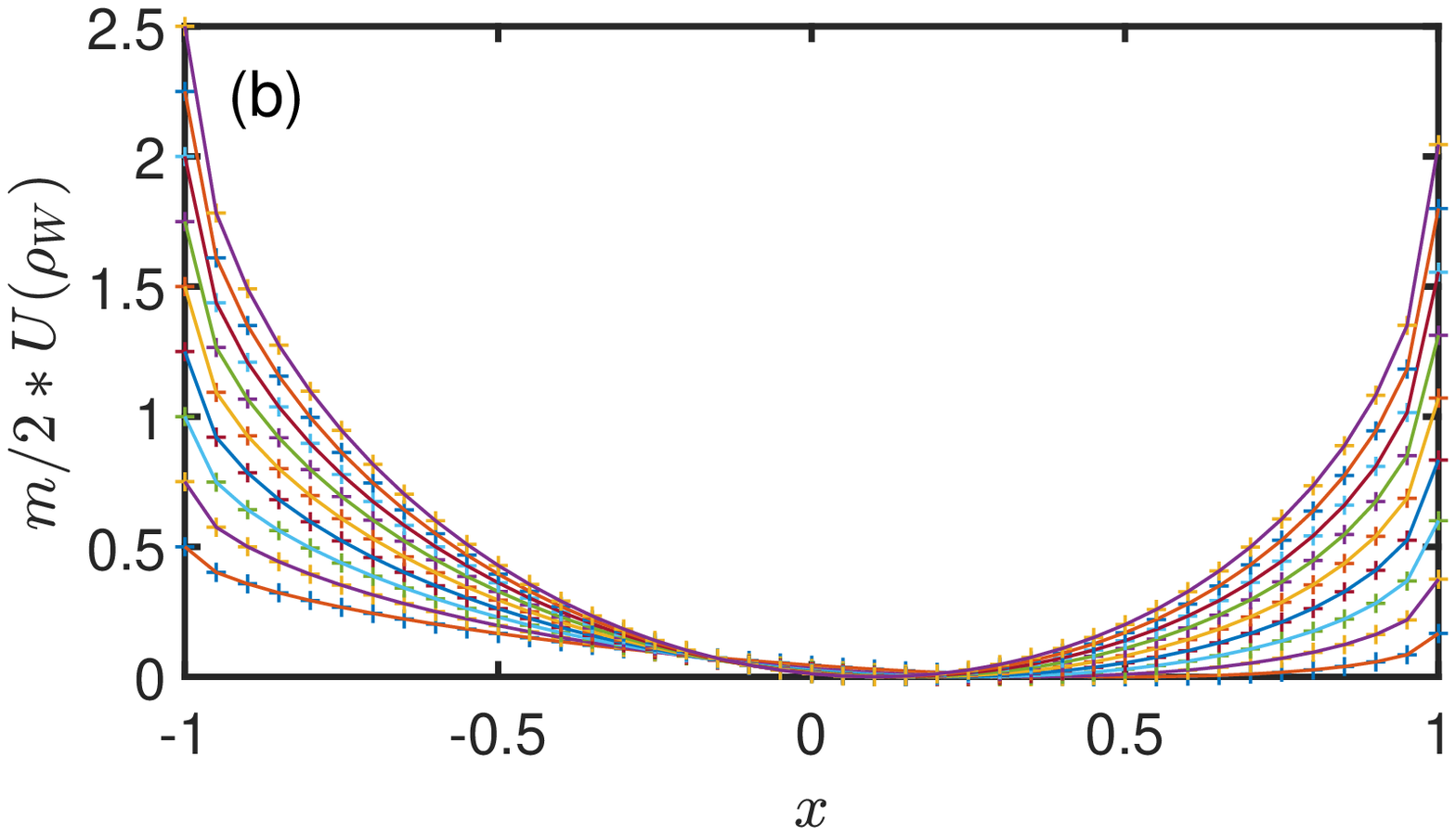}}%
\newline
\caption{(color online) The measurement-induced non-locality $U(\protect\rho _{I})$ for the Isotropic
states (a) and the Werner states (b) vs $x$. The solid line corresponds to
the strictly analytical expression of Eq. (\protect\ref{isotropic}) for (a)
and Eq. (\protect\ref{Werner}) for (b), while the numerical expressions of
Eq. (\protect\ref{theorem}) in both figures are marked by the "+" . }
\end{figure}

Before the next example, we would like to emphasize that in both example
(iv) and example (v), the reduced density matrices $\rho_A$ is the maximally
mixed state which implies the complete degeneracy. Intuitively, this belongs
to the worst case mentioned in the last section (it requires the
optimization in the total space of subsystem A). However, in the analytic
procedure, one can find that all that are needed to be optimized can be
automatically eliminated, which means no optimization is practically
required. But it doesn't mean that the optimization isn't performed in the
numerical procedure. One can find that our numerical method stably
approaches the unique optimal solution given in the analytic procedure. In
this sense, the two examples provide more powerful proofs for the
effectiveness of our theorem than other examples. In addition, considering
the dual definitions of the measurement-induced non-locality and LQU, one can find that $U=Q$ for both the
isotropic states and the Werner states, since no optimization is practically
covered, which is consistent with Ref. \cite{remed}.

\textit{(vi) The $(m\otimes m)$-dimensional degenerate and non-degenerate
hybrid states.-} In order to demonstrate the practicability of our theorem
and avoid the uniform degeneracy of the reduced density matrix, we construct
a particular state as
\begin{equation}
\rho _{H}=\frac{\rho _{W}+P}{2}+\frac{(1-m+x)}{2(m^{2}-1)}\left(
P-\left\vert 11\right\rangle \left\langle 11\right\vert \right) ,x\in
\lbrack -1,1],
\end{equation}%
with $P=\frac{1}{n}\sum_{k=2}^{n+1}\left\vert kk\right\rangle \left\langle
kk\right\vert $ and generally $m\geq 3$ and $n\in \lbrack 2,m-1]$ to be
supposed. The reduced density matrix $\rho _{A}$ of this state has three
different eigenvalues. Among them, there could be two degenerate subspaces
and one non-degenerate subspace. The eigenvalue $\frac{2m^{2}-mx-m-1}{%
2m(m^{2}-1)}$ is non-degenerate, the eigenvalue $\frac{m^{3}+mx-1}{%
2m(m^{2}-1)}$ is $n$-fold degenerate and the eigenvalue $\frac{1}{2m}$ is $%
(m-n-1)$-fold degenerate. However, one can note that if $m=3$, then $n=2$.
Thus $(m-n-1)=0$, i.e., this subspace corresponding to the eigenvalue $\frac{%
1}{2m}$ doesn't exist, which implies that the state has two different
eigenvalues with only one degenerate subspace. When $m=3$, one can find that
the measurement-induced non-locality $U(\rho _{H})$ based on our definition is%
\begin{equation}
U(\rho _{H})=\frac{9-3x-6\sqrt{2-2x^{2}}+(2-t)f(x)}{48},  \label{Uhr}
\end{equation}%
where%
\begin{equation}
t=\left\{
\begin{array}{cc}
m, & -1<x\leq -\frac{14}{15} \\
1, & -\frac{14}{15}<x\leq 1%
\end{array}%
\right. ,
\end{equation}%
and
\begin{eqnarray}
&&f(x)=11+3.5x-\sqrt{(7x+22)(x+1)}  \notag \\
&&+2\sqrt{2-2x^{2}}-\sqrt{(4x+44)(1-x)}.
\end{eqnarray}%
When $m>3$, the measurement-induced non-locality $U(\rho _{H})$ can be given by
\begin{eqnarray}
&&U(\rho _{H})=\frac{3}{4}-\frac{1}{4(m^{3}-m)}[2(m^{2}-x-1)  \notag \\
&&+X_{+}(m-2n)+2Y+(m^{2}-m-2n+2)\sqrt{X_{+}X_{-}}  \notag \\
&&+(2n-2)\sqrt{X_{+}+Y}(\sqrt{X_{+}}+\sqrt{X_{-}})],  \label{Uhri}
\end{eqnarray}%
where $X_{\pm }=(m\mp 1)(1\pm x)$ and $Y=(m^{2}+x-m)\frac{m}{n}$.
\begin{figure}[tbp]
\centering
\includegraphics[width=1\columnwidth,height=2in]{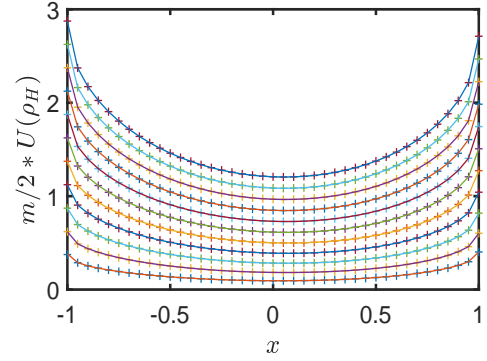}
\caption{(color online) The measurement-induced non-locality $U(\protect\rho _{H})$ for hybrid states vs $%
x$. The solid line and the "+" line correspond to the strictly analytical
expression of Eq. (\protect\ref{Uhr}), Eq. (\protect\ref{Uhri}) and the
numerical expression of Eq. (\protect\ref{theorem}). }
\end{figure}
The comparison of the numerical and the analytic expressions is given in
FIG. 4 which again shows the perfect consistency.

\textit{(vii) The general quantum states.} In order to show the power of our
theorem for high-dimensional states, we consider such a state $\rho_G$ by
mixing a maximally mixed state and a randomly generated $(400\times 400)$%
-dimensional mixed state $G$ as
\begin{equation}
\rho _{G}=\frac{(1-x)}{400}\mathbb{I}_{400}+xG, x\in [ 0,1].
\end{equation}%
Sine the matrix $G$ is too large, it is impossible to explicitly give here.
The measurement-induced non-locality versus $x$ has been plotted in FIG. 5. We have uniformly take $51$
points in $x\in [0,1]$ and it averagely takes about $8.4$ seconds for the
program (Matlab R2014b) running on the personal laptop (2.8 GHz Intel Core
i7/16 GB 1600 MHz DDR3).
\begin{figure}[tbp]
\centering
\includegraphics[width=1\columnwidth,height=2in]{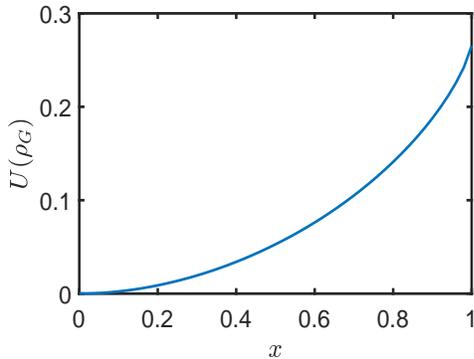}
\caption{(color online) The measurement-induced non-locality $U(\protect\rho _{G})$ for general hybrid
states vs $x$. The solid line corresponds to the numerical expression of
Eq. (\protect\ref{theorem}).}
\end{figure}

\section{Conclusions and discussion}

We have redefined the measurement-induced non-locality based on the skew information associated with the
broken observable instead of the original high-rank observable. The new
definition can solve the non-contractivity problem in the previous measurement-induced non-locality based
on the $l_2$ norm and has obvious operational meaning in terms of quantum
metrology. It allows us to analytically calculate the measurement-induced non-locality of the pure
states, the qubit-qudit states, and some large classes of high-dimensional
states. In particular, the new definition enables us to develop the powerful
inverse approximate joint diagonalization algorithm based on the very remarkable Jacobi method for the approximate joint diagonalization
problem. It is shown that this inverse approximate joint diagonalization algorithm similar to the corresponding
approximate joint diagonalization algorithm has good effectiveness such as simplicity, stability, high
efficiency and state independence. This is further proven by the detailed
comparisons between the analytic and the numerical results of various
examples. Compared with the diagonalization of a single density matrix, it
is shown that our measurement-induced non-locality has even the almost analytic expression for any
quantum state, which gives an alternative effective means for the
computability of the measurement-induced non-locality.

Finally, we would like to emphasize that the method to breaking observable
isn't a trivial skill but has many successful applications and can conquer
many key problems in quantum discord and especially in quantum coherence.
The potential application is worth our forthcoming attention. In addition,
considering the quantum resource theory, it could be usually impossible for
a general state to find an analytic quantifier of a general quantum
resource, so how to develop an effective numerical means should be a quite
necessary problem. The power exhibited by the inverse approximate joint diagonalization and the approximate joint diagonalization algorithms could
shed new light on both the resource theory and the other aspects of physics.

\acknowledgments This work was supported by the National Natural Science
Foundation of China, under Grant No.11775040 and 11375036, and the Xinghai
Scholar Cultivation Plan.


\begin{thebibliography}{99}
\bibitem{B1} C.H. Bennett, H.J. Bernstein, S. Popescu, and B. Schumacher,
Concentrating partial entanglement by local operations, Phys. Rev. A \textbf{%
53}, 2046 (1996).

\bibitem{B2} C. H. Bennett, D. P. DiVincenzo, J. A. Smolin, and W. K.
Wootters, Mixed-state entanglement and quantum error correction, Phys. Rev.
A \textbf{54}, 3824 (1996).

\bibitem{B3} C. H. Bennett, G. Brassard, S. Popescu, B. Schumacher, J. A.
Smolin, and W. K. Wootters, Purification of Noisy Entanglement and Faithful
Teleportation via Noisy Channels, Phys. Rev. Lett. \textbf{76}, 722 (1996).

\bibitem{o1} P. Horodecki, R. Horodecki, Distillation and bound
entanglement, Quant. Inf. Comp. \textbf{1}, 45 (2000).

\bibitem{o2} W. K. Wootters, Entanglement of formation and concurrence,
Quant. Inf. Comp. \textbf{81}, 865 (2009).

\bibitem{o3} R. Horodecki, P. Horodecki, M. Horodecki and K. Horodecki,
Quantum entanglement, Rev. Mod. Phys. \textbf{81}, 865 (2009).

\bibitem{d1} L. Henderson, V. Vedral, Classical, quantum and total
correlations, J. Phys. A, Math. Theor. \textbf{34} 6899 (2001).

\bibitem{d2} H. Ollivier, W. H. Zurek, Quantum discord: A measure of the
quantumness of correlations, Phys. Rev. Lett. \textbf{88} 017901 (2001).

\bibitem{d3} A. Datta, A. Shaji, and C. M. Caves, Quantum discord and the
power of one qubit, Phys. Rev. Lett. \textbf{100}, 050502 (2008).

\bibitem{d4} T. Tufarelli, D. Girolami, R. Vasile, S. Bose, and G. Adesso,
Quantum resources for hybrid communication via qubit-oscillator states,
Phys. Rev. A \textbf{86}, 052326 (2012).

\bibitem{d5} C. S. Yu, J. Zhang, and H. Fan, Quantum dissonance is rejected
in an overlap measurement scheme, Phys. Rev. A \textbf{86}, 052317 (2012).

\bibitem{d6} A. Brodutch, Discord and quantum computational resources, Phys.
Rev. A \textbf{88}, 022307 (2013).

\bibitem{c1} T. Baumgratz, M. Cramer, and M. B. Plenio, Quantifying
coherence, Phys. Rev. Lett. \textbf{113}, 140401 (2014).

\bibitem{c2} A. Winter and D. Yang, Operational resource theory of
coherence, Phys. Rev. Lett. \textbf{116}, 120404 (2016).

\bibitem{c3} E. Chitambar and M. H. Hsieh, Relating the resource theories of
entanglement and quantum coherence, Phys. Rev. Lett. \textbf{117}, 020402
(2016).

\bibitem{c4} A. Streltsov, S. Rana, P. Boes and J. Eisert, Structure of the
resource theory of quantum coherence, Phys. Rev. Lett. \textbf{119}, 140402
(2017).

\bibitem{c5} K. B. Dana, M. G. D\'{\i}az, M. Mejatty and A. Winter, Resource
theory of coherence: beyond states, Phys. Rev. A \textbf{95}, 062327 (2017).

\bibitem{c6} A. Streltsov, S. Rana, M. N. Bera and M. Lewenstein, Towards
resource theory of coherence in distributed scenarios, Phys. Rev. X \textbf{7%
}, 011024 (2017).

\bibitem{t1} F. G. S. L. Brand\~{a}o, M. Horodecki, J. Oppenheim, J. M.
Renes, and R. W. Spekkens, Resource theory of quantum states out of thermal
equilibrium, Phys. Rev. Lett. \textbf{111}, 250404 (2013).

\bibitem{t2} F. G. S. L. Brand\~{a}o and G. Gour, Reversible framework for
quantum resource theories, Phys. Rev. Lett. \textbf{115}, 070503 (2015).

\bibitem{yuc} C. S. Yu, Quantum coherence via skew information and its
polygamy, Phys. Rev. A \textbf{95}, 042337 (2017).

\bibitem{peres} A. Peres, Separability criterion for density matrices, Phys.
Rev. Lett. \textbf{77}, 1413 (1996).

\bibitem{wootters} W. K. Wootters, Entanglement of formation of an arbitrary
state of two qubits, Phys. Rev. Lett. \textbf{80}, 2245 (1998).

\bibitem{vidal} G. Vidal, and R. F. Werner, Computable measure of
entanglement, Phys. Rev. A \textbf{65}, 032314 (2002).

\bibitem{geom1} S. L. Luo, Quantum discord for two-qubit systems, Phys. Rev.
A \textbf{77}, 042303 (2008).

\bibitem{geom} S. L. Luo, Using measurement-induced disturbance to
characterize correlations as classical or quantum, Phys. Rev. A \textbf{77},
022301 (2008).


\bibitem{dak} B. Daki\'c, V. Vedral, and \v{C}. Brukner, Necessary and
sufficient condition for nonzero quantum discord, Phys. Rev. Lett. \textbf{%
105}, 190502 (2010).

\bibitem{giro} D. Girolami, T. Tufarelli, and G. Adesso, Characterizing
nonclassical correlations via local quantum uncertainty, Phys. Rev. Lett.
\textbf{110}, 240402 (2013).

\bibitem{diagon} C. S. Yu, S. X. Wu, X. G. Wang, X. X. Yi and H. S. Song,
Quantum correlation measure in arbitrary bipartite systems, Europhys. Lett.
\textbf{107}, 10007 (2017).

\bibitem{florian} F. Mintert, M. Ku\'s, and A. Buchleitner, Concurrence of
mixed bipartite quantum states in arbitrary dimensions, Phys. Rev. Lett.
\textbf{92}, 167902 (2004).

\bibitem{kai} K. Chen, S. Albeverio and S. M. Fei, Concurrence of arbitrary
dimensional bipartite quantum states, Phys. Rev. Lett. \textbf{95}, 040504
(2005).

\bibitem{ycc} C. S. Yu, and H. S. Song, Separability criterion of tripartite
qubit systems, Phys. Rev. A \textbf{72}, 022333 (2005).

\bibitem{nc} K. Audenaert, F. Verstraete, and B. D. Moor, Variational
characterizations of separability and entanglement of formation, Phys. Rev.
A \textbf{64}, 052304 (2001).

\bibitem{semi} M. A. Jafarizadenh, M. Mirzaee, and M. Rezaee, Best separable
approximation with semi-definite programming method, Int. J. Quant. Inf.
\textbf{2}, 541 (2004).

\bibitem{nr} J. \v{R}eh\'a\v{c}ek and Z. Hradil, Quantification of
entanglement by means of convergent iterations, Phys. Rev. Lett. \textbf{90}%
, 127904 (2003).

\bibitem{nume} Y. Zinchenko, S. Friedland, and G. Gour, Numerical estimation
of the relative entropy of entanglement, Phys. Rev. A \textbf{82}, 052336
(2010).

\bibitem{nm} K. Cao, Z. W. Zhou, G. C. Guo and L. X. He, Efficient numerical
method to calculate the three-tangle of mixed states, Phys. Rev. A \textbf{81%
}, 034302 (2010).

\bibitem{compu} B. R\"{o}thlisberger, J. Lehmann and D. Loss, Comp. Phys.
Commu. \textbf{183}, 155 (2012).

\bibitem{highly} B. R\"{o}thlisberger, J. Lehmann, D. S. Saraga, P. Traber
and D. Loss, Highly entangled ground states in tripartite qubit systems,
Phys. Rev. Lett. \textbf{100}, 100502 (2008).

\bibitem{numerical} B. R\"{o}thlisberger, J. Lehmann and D. Loss, Numerical
evaluation of convex-roof entanglement measures with applications to spin
rings, Phys. Rev. A \textbf{80}, 042301 (2009).

\bibitem{robu} C. Napoli, T. R. Bromley, M. Cianciaruso, M. Piani, N.
Johnston and G. Adesso, Robustness of coherence: an operational and
observable measure of quantum coherence, Phys. Rev. Lett. \textbf{116},
150502 (2016).

\bibitem{asym} M. Piani, M. Cianciaruso, T. R. Bromley, C. Napoli, N.
Johnston and G. Adesso, Robustness of asymmetry and coherence of quantum
states, Phys. Rev. A \textbf{93}, 042107 (2016).

\bibitem{trac} S. Rana, P. Parashar and M. Lewenstein, Trace-distance
measure of coherence, Phys. Rev. A \textbf{93}, 012110 (2016).

\bibitem{gene} P. Zanardi, G. Styliaris and L. C. Venuti, Measures of
coherence-generating power for quantum unital operations, Phys. Rev. Lett.
\textbf{95}, 052307 (2017).

\bibitem{loca} S. L. Luo and S. S. Fu, Measurement-Induced Nonlocality,
Phys. Rev. Lett. \textbf{106}, 120401 (2011).

\bibitem{geometr} S. L. Luo, and S. S. Fu, Geometric measure of quantum
discord, Phys. Rev. A \textbf{82}, 034302 (2010).

\bibitem{piani} M. Piani, Problem with geometric discord, Phys. Rev. A
\textbf{86}, 034101 (2012).

\bibitem{SY} S. Y. Mirafzali, I. Sargolzahi, A . Ahanj, K. Javidan, and M.
Sarbishaei, Measurement-induced nonlocality for an arbitrary bipartite
state, Quant. Inf. Comp. \textbf{13}, 0479 (2011).

\bibitem{gamy} A. Sen, D. Sarkar, and A. Bhar, Monogamy of
measurement-induced nonlocality, J. Phys. A Math. Theor., \textbf{45},
405306 (2012).

\bibitem{Rana} S. Rana, and P. Parashar, Geometric discord and
measurement-induced nonlocality for well known bound entangled states,
Quant. Inf. Proc. \textbf{12}, 2523 (2013).

\bibitem{Wu} S. X. Wu, J. Zhang, C. S. Yu, and H. S. Song,
Uncertainty-induced quantum nonlocality, Phys. Letts. A \textbf{378}, 344
(2014).

\bibitem{Hu} M. L. Hu, and H. Fan, Measurement-induced nonlocality based on
the trace norm, New. J. Phys. \textbf{17},033004 (2015).

\bibitem{maximal} L. D. Wang, L. T. Wang, M. Yang, J. Z. Xu, Z. D. Wang, and
Y. K. Bai, Entanglement and measurement-induced nonlocality of mixed
maximally entangled states in multipartite dynamics, Phys. Rev. A \textbf{93}%
, 062309 (2016).

\bibitem{based} Z. J. Xi, X. G. Wang and Y. M. Li, Measurement-induced
nonlocality based on the relative entropy, Phys. Rev. A \textbf{85}, 042325
(2012).

\bibitem{skew} E. P. Wiger and M. M. Yanase, Information contents of
distributions, Proc. Natl. Acad. Sci. USA, \textbf{49}, 910 (1963).

\bibitem{convex} E. H. Lieb, Convex trace functions and the
Wigner-Yanase-Dyson conjecture, Adv. Math. \textbf{11}, 267 (1973).

\bibitem{uncer} S. L. Luo, Wigner-Yanase skew information and uncertainty
relations, Phys. Rev. Lett. \textbf{91}, 180403 (2003).

\bibitem{fis2} U. Dorner, R. Demkowicz-Dobrzanski, B. J. Smith, J. S.
Lundeen, W.Wasilewski, K. Banaszek, and I. A. Walmsley, Phys. Rev. Lett.
\textbf{102}, 040403 (2009).

\bibitem{fis1} S. L. Braunstein and C. M. Caves, Phys. Rev. Lett. \textbf{72}%
, 3439 (1994);

\bibitem{fis11} S. L. Braunstein, C. M. Caves, and G. J. Milburn, Ann. Phys.
(N.Y.)\textbf{247}, 135 (1996).

\bibitem{luo1} S. L. Luo, Proc. Am. Math. Soc. \textbf{132}, 885 (2003).

\bibitem{blind} J. F. Cordoso, and A. Souloumiac, Blind beamforming for
non-Gaussian signals, Radar \& Signal Processing, IEE Proceedings-F \textbf{%
140}, 362 (1993).

\bibitem{jacibo} J. F. Cordoso, and A. Souloumiac, Jacobo angles for
simultaneous diagonalization, SIAM J. Mater. Anal. Appl. \textbf{17}, 161
(1996).

\bibitem{fast} A. Ziehe, P. Laskov, G. Nolte, and K.R. M\"{u}ller, A fast
algorithm for joint diagonalization with non-orthogonal transformations and
its application to blind source separation, J. Mach. Learn. Res. \textbf{5},
777 (2004).

\bibitem{golub} G. H. Golub, and C. F. Van Loan, \textit{Matrix Computations}%
, 3rd edition (The Johns Hopkins University Press, 1996).

\bibitem{vanloanbook} C. F. Van Loan, and N. P. Pitsianis, \textit{Linear
algebra for large scale and real time applications}, edited by M. S. Moonen
and G. H. Golub, (Kluwer, Dordrecht, pp. 293--314, 1993).

\bibitem{Alber} G. Alber, T. Beth, M. Horodecki, P. Horodecki, R. Horodecki,
M. R\"otteler, H. Weinfurter, R. Werner, and A. Zeilinger, \textit{Quantum
Information: An Introduction to Basic Theoretical Concepts and Experiments},
(Springer-Verlag, Berlin, Heidelberg, 2001).

\bibitem{remed} L. N. Chang, and S. L. Luo, Remedying the local ancilla
problem with geometric discord, Phys. Rev. A \textbf{87}, 062303 (2013).
\end{thebibliography}
\end{document}